\documentclass[10pt,letterpaper]{article}
\usepackage{opex3}
\usepackage{cite}

\begin{document}

\title{Propagation of plasmons in designed single crystalline silver nanostructures}

\author{Shailesh Kumar,$^{*}$ Ying-Wei Lu, Alexander Huck, and Ulrik L. Andersen}

\address{Department of Physics, Technical University of Denmark, Building 309, 2800 Lyngby, Denmark}

\email{$^{*}$shailesh@fysik.dtu.dk} 


\begin{abstract*}
We demonstrate  propagation of plasmons in single crystalline silver nanostructures fabricated using a combination of a bottom-up and a top-down approach. Silver nanoplates of thickness around 65~nm and a surface area of about $100~\mu m^2$ are made using a wet chemical method. Silver nanotips and nanowires are then sculptured by focused ion beam milling. The plasmons are excited by using the fluorescence from the redeposited silver clusters during the milling process. Propagation of plasmons in the nanowires is observed in the visible spectral region. We also observe a cavity effect by measuring the emission spectrum from the distal wire end.
\end{abstract*}



\section{Introduction}

Optical fields supported by metallic nanostructures - the surface plasmons - can be tightly confined to transverse dimensions much smaller than the wavelength of the associated light field. Such a result has great potential for the miniaturization of photonic circuits~\cite{2010Bozhevolnyi, 2010Schuller} and for enabling strong coherent interaction of individual quantum emitters with single mode spatial optical fields~\cite{2006Chang}, which eventually offer chip based realizations of quantum information processing~\cite{2007Chang}, high-sensitive sensing and high-speed optical communication~\cite{2010Bozhevolnyi, 2010Schuller}.

However, as a result of the strong field confinement and thus field enhancement, the requirements on the fabrication of the nanostructures become very  critical. Nanometer sized defects in the fabrication process may result in unwanted scattering centers leading to poor optical performance in terms of increased optical losses~\cite{PRLAgCavity} and reduced field confinement~\cite{2010Hecht}. Therefore, to circumvent such detrimental effects, it is important to develop a fabrication method that produces single-crystalline structures with atomically smooth surfaces. 

Single-crystalline nanowires of silver and gold with exquisite propagation properties have been fabricated using chemical synthesis~\cite{Auwirefabrication, Agwirefabrication} and have been successfully employed to harvest single photons~\cite{2007Akimov,2009Kolesov,2011Huck}. The drawback, however, of this fabrication method is that predesigned structures cannot be made~\cite{Auplatefabrication, MicrosizedAgcrystals}. To develop nanocircuitry for controlled light guiding and enhanced coupling, another method must be considered. The standard approach for obtaining the desired metallic nanostructures has been electron-beam lithography followed by thermal evaporation of metal~\cite{Refereedemand}. This method, however, results in polycrystalline metallic nanostructures which leads to severe scattering of a propagating field from the grain boundaries~\cite{PRLAgCavity}. Thermal evaporation of metal is thus unsuitable for the fabrication of low loss metallic nanocircuitry.    

Recently, a different and very promising method that combines bottom-up and top-down approaches has been applied to fabricate complex, single-crystalline nanostructures of gold~\cite{2010Hecht, Goldflakemachine}. Single-crystalline gold plates were produced using chemical synthesis~\cite{Auplatefabrication} and, subsequently, the desired structure was formed by the use of focused ion beam (FIB) milling. Using this method, they showed a dramatic increase in the surface smoothness of some gold nanoantennas. However, due to the high intrinsic losses of gold, the losses in these structures remained large. These losses are in particular large in the visible and near-infrared optical spectrum, due to the relatively low plasma frequency and high interband transition strengths of gold~\cite{JohnsonChristy, Lowlossmaterial}. On the other hand, the intrinsic loss of silver is much lower and is therefore a more natural choice in many applications. 
 
In this paper, we demonstrate the propagation of plasmons in predesigned single-crystalline silver nanowires fabricated using chemical synthesis of silver nanoplates (SNPs) and FIB milling of them. We excite the silver clusters redeposited in the FIB milling process, from which the fluorescence couples to the propagating mode of the plasmonic structure. The cavity effect in the nanowires is also observed. We make nanowires and nanotips with prespecified dimensions and with a high degree of smoothness.

\section{Chemical synthesis of SNPs}

The first step in producing single-crystalline and designed silver nanostructures is the fabrication of SNPs. Various chemical procedures have been followed to synthesize such SNPs~\cite{MicrosizedAgcrystals,Liquidliquidinterface,langmuir,solvothermal}. In all these previous accounts, however, the SNPs were either too small~\cite{solvothermal} or too thick~\cite{MicrosizedAgcrystals}, to allow for the carving of nanoplasmonic circuits from them. We synthesize single crystalline SNPs of average thickness of around 65~nm and with areas of around $100~\mu {m}^2$ by modification of a process~\cite{MicrosizedAgcrystals}, which was previously used to synthesize silver plates of thicknesses of about $0.5~\mu m$. All the chemicals used to make the SNPs were purchased from Sigma-Aldrich. Our chemical process starts by preparation of solutions of 100~mg silver nitrate ($AgNO_3$) in 30~ml deionized (DI) water, 15~mg of iron sulfate heptahydrate (${FeSO}_4 \cdot {7H}_2{O}$) in 30~ml DI water, and 200~mg of  polyvinyl pyrollidone ($PVP-K30$) in 40~ml of DI water. These solutions were cooled to temperatures between 8 and 20 degrees celcius. The silver nitrate solution was then put into a conical flask at room temperature and 3.6~mg of ${H}_2 {SO}_4$ was added to the solution. Then the $PVP-K30$ solution was added to the mixture, followed by addition of the ${FeSO}_4 \cdot {7H}_2{O}$ solution. The reaction was run for 3 hours at room temperature under continuous stirring using a magnetic stirrer. Then the solution was centrifuged at 4000~rpm for half an hour, and SNPs were obtained as precipitate. The SNPs thus obtained were washed twice with ethanol and once with water. After washing, the SNPs were stored dissolved in water. We also made SNPs using different weights of ${FeSO}_4 \cdot {7H}_2{O}$, where 50~mg, 100~mg and 150~mg of ${FeSO}_4 \cdot {7H}_2{O}$ (referred to as processes 2, 3 and 4, respectively, in the following discussion) was used instead of 15~mg of ${FeSO}_4 \cdot {7H}_2{O}$ (referred to as process 1 in the following discussion).

\section{Characterization of SNPs}

\begin{figure}[htbp]
\begin{center}
\includegraphics[]{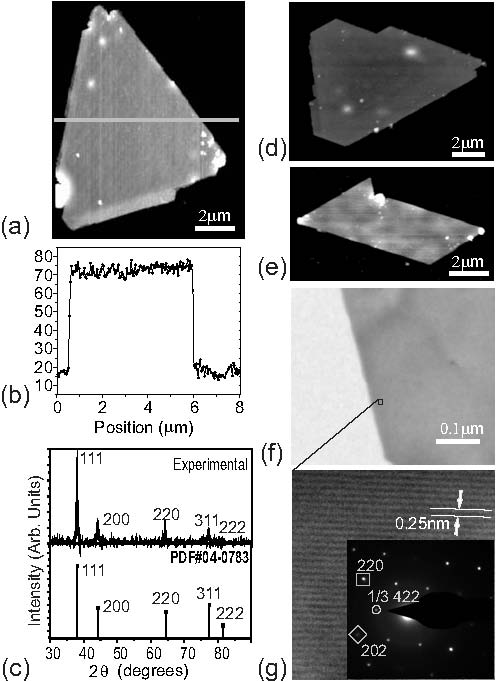}
\caption[Characterization of silver nanoplates.]{ \textbf{Structural characterization of SNPs}  \textbf{(a),(d)} and \textbf{(e)} AFM images of SNPs obtained with process 1.   \textbf{(b)} Height profile of the SNP shown in (a) along the horizontal line indicated in (a).  \textbf{(c)} X-ray diffraction pattern for the SNPs along with standard pattern for silver (PDF$\#04-0783$), showing dominance of the (111) crystalline plane for SNPs. \textbf{(f)} TEM image showing the edge of an SNP. \textbf{(g)} High resolution TEM image of the SNP from which the lattice spacing is clearly visible. The inset shows a selected area electron diffraction pattern, which proves the single crystallinity of our SNPs. \label{FIB:AFM_images}}
\end{center}
\end{figure}

To characterize the dimensions of the SNPs thus obtained, the solutions containing SNPs were spin coated on a fused silica substrate, and AFM images of SNPs were taken. AFM images of SNPs made using process 1 are shown in Fig.~\ref{FIB:AFM_images}(a),~\ref{FIB:AFM_images}(d) and~\ref{FIB:AFM_images}(e). The number of SNPs on the substrate in an area of $100 \times 100~ \mu {m}^2$ was 1 or 2. From the AFM images, it can be seen that the solutions contain some silver particles in addition to the SNPs, which are side products obtained in small amount. It was also observed that the SNPs obtained with different processes were of uniform thickness. Height profile of an SNP along the line shown in Fig.~\ref{FIB:AFM_images}(a) is presented in Fig.~\ref{FIB:AFM_images}(b). We have also conducted X-ray diffraction and transmission electron microscope (TEM) studies, which confirmed the single crystallinity of our SNPs, summarized in Fig.~\ref{FIB:AFM_images}(c),~\ref{FIB:AFM_images}(f) and~\ref{FIB:AFM_images}(g). In table~\ref{table:Chemical_processes}, we present the average thickness and the average maximum lengths of the SNPs obtained using different synthesization processes. The root mean square (RMS) roughness in table~\ref{table:Chemical_processes}, represents the variation in the thickness over an area of $1~\mu {m}^2$. To obtain the average RMS roughness of SNPs independent on the substrate, we first made measurements of the roughness of the substrate and subsequently subtracted this value from the value obtained by measuring directly onto the SNPs. This then gave us a good estimate of the roughness stemming directly from the SNPs. From table~\ref{table:Chemical_processes}, it is clear that SNPs obtained in process 1 are thinner, have larger transverse dimensions and are smoother than SNPs obtained with the other processes. Therefore, SNPs obtained in process 1 are more suitable for fabrication of miniaturized plasmonic circuits. 

\begin{table}
\caption{Properties of SNPs obtained using different chemical processes.}
\begin{center}
\begin{tabular}{| c | c | c | c |c | c |}
  \hline                        
  	&  Average &  Average  & Average RMS &  Average RMS &  Average  RMS\\
Process & thickness, &   maximum  &  roughness, &  roughness, SNP  & roughness, \\
	& SNP &  lengths, SNP & Substrate & on substrate & SNP\\
 & (nm) & ($\mu$m) & (nm) & (nm) & (nm)\\
\hline

1 & 63.6 & 10.75 & 3.87  & 4.78 & 0.91\\

2&97.2&8.51&4.01& 6.76&2.75\\

3&151.7&8.59&4.43&7.12&2.69\\

4 & 259.0 & 3.13 & 4.14 & 5.99 & 1.85\\
\hline  
\end{tabular}
\end{center}
\label{table:Chemical_processes}
\end{table}

\section{Fabrication of designed structures using focused ion beam lithography}

\begin{figure}[htbp]
\begin{center}
\includegraphics[]{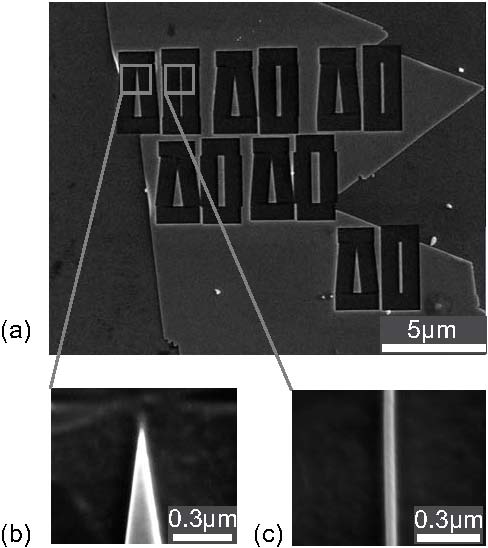}
\caption[SEM images of structures obtained from FIB milling of silver nanoplates.]{\textbf{SEM images of structures made with FIB milling of an SNP.} \textbf{(a)} Some tips and nanowires of different tip angles and wire widths made using FIB milling of an SNP.  \textbf{(b)} Magnified image of tip shown in left rectangle in (a).  \textbf{(c)} Magnified image of wire shown in right rectangle in (a).  \label{FIB_structures}}
\end{center}
\end{figure}

The desired structures were fabricated by cutting the SNPs by means of FIB milling (Helios Nanolab from FEI). We made silver wires and tips, which have been found to be effective for coupling quantum emitters to propagating plasmonic  modes~\cite{2006Chang}. The substrate used for FIB milling of SNPs was a fused silica glass plate coated with a 50~nm layer of conducting Indium Tin Oxide (ITO), which is required for FIB milling. We plasma cleaned our substrates for 5~minutes. Then, the solution containing SNPs was spin coated onto the substrates. To obtain nanostructures in the FIB milling process, we chose an acceleration voltage of 30~kV and a Ga-ion current of 1.5~pA. Figure~\ref{FIB_structures}(a) shows  a  scanning electron microscope (SEM) image of some wires and tip structures made with FIB milling of a large SNP, whereas Fig.~\ref{FIB_structures}(b) and~\ref{FIB_structures}(c) show zoom-in high resolution SEM images. The apex angle of the tip shown is $10\,^{\circ}$, and width of the wire shown is 50~nm. Structures obtained with this method are smooth (roughness of the order of the SNP roughness) and single crystalline.

\section{Optical characterization of structures made with FIB milling}

When SNPs are FIB milled with Ga ions, very small clusters of silver gets redeposited on silver structures~\cite{FIBbook}. The silver clusters have sizes less than the resolution of SEM images taken ($\approx 2~{nm}$), and can be as small as a cluster of few atoms. These small clusters of silver give rise to fluorescence~\cite{Agfluorescence} when illuminated by light. Another source of fluorescence could be some Ga ions  deposited during the FIB process. This is however less likely as the  concentration of Ga ions deposited is very low~\cite{2010Hecht}.
               
\begin{figure}[htbp]
\begin{center}
\includegraphics[]{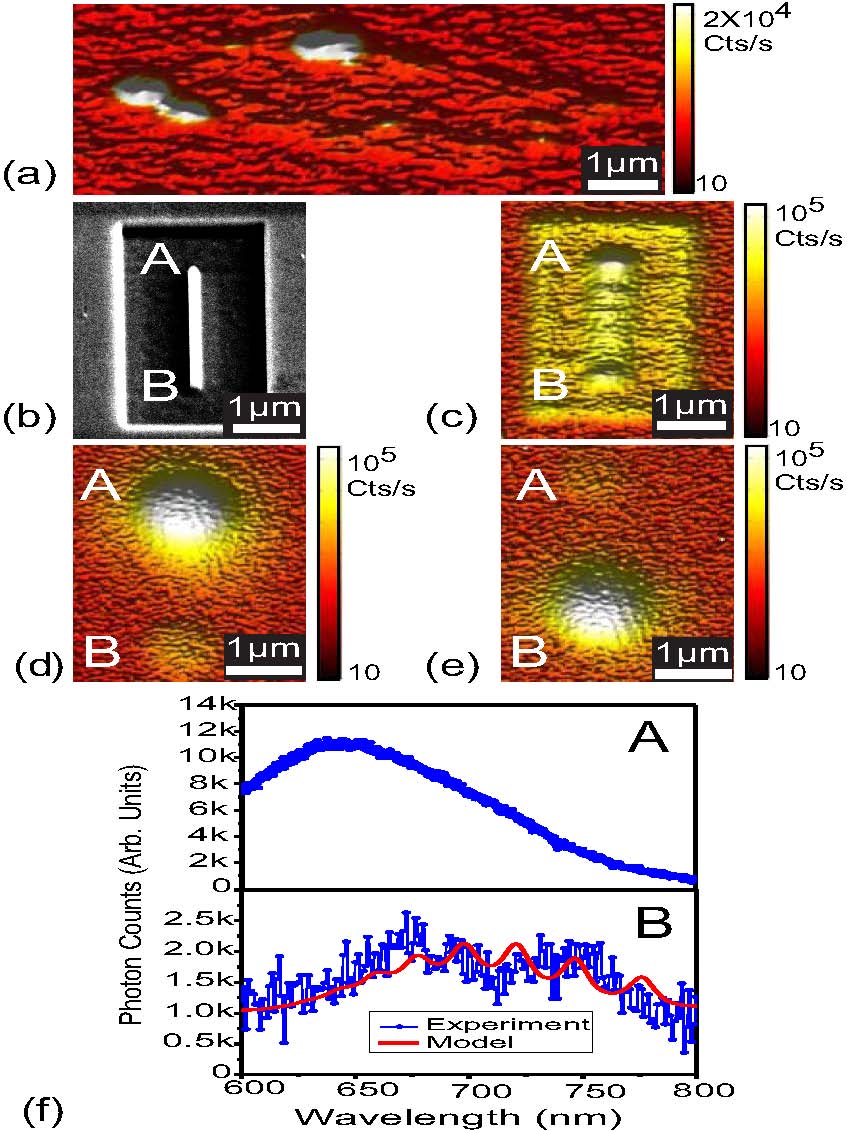}
\caption[Propagation of plasmons in FIB milled silver nanowire.]{\textbf{Optical Characterization.} \textbf{(a)} Fluorescence image of the SNP shown in Fig. 1(e). Color scale for all the fluorescence images is exponential. \textbf{(b)} SEM image of a wire made by FIB milling of an SNP.  \textbf{(c)} Confocal microscope fluorescence image of the area shown in (b).   \textbf{(d)}  and \textbf{(e)} Galvanometric fluorescence images taken when the laser is focused onto ends \lq A\rq ~and \lq B\rq ~of the wire shown in (c), respectively.  \textbf{(f)} Spectrum taken at  end \lq A\rq ~and end \lq B\rq ~of the wire in  (d). A model based on the spectrum of end \lq A\rq ~as input and silver wire as a Fabry-Perot cavity is also plotted. \label{optical_characterization}}
\end{center}
\end{figure}

To optically characterize the structures, we used a standard home built fluorescence confocal microscope in which green light at 532~nm is focused onto the structures, and the fluorescence light is collected by a high numerical aperture (0.95) lens and directed through filters (transmits the wavelength range from 647~nm to 785~nm) to an avalanche photodiode (APD). The focal spot for the excitation laser has a diameter of around 300~nm. A fluorescence image of an SNP taken at an excitation power of $80~\mu W$ (before FIB milling) is shown in Fig.~\ref{optical_characterization}(a),  where it can be observed that the fluorescence from SNP is negligible. By comparing the fluorescence image in Fig.~\ref{optical_characterization}(a) to SEM image of the SNP in Fig.~\ref{FIB:AFM_images}(e), we conclude that the fluorescence stems from particles lying close to the SNP. A fluorescence image of the wire taken with an excitation power of $8~\mu W$ and polarization perpendicular to the long axis of the wire, whose SEM image is shown in Fig.~\ref{optical_characterization}(b), is shown in Fig.~\ref{optical_characterization}(c). The wire has a length of $3~\mu m$, a width of 100~nm and a height of approximately 100~nm. It is clearly seen that there is pronounced fluorescence which presumably stems from the clusters of silver particles. 
 
We now demonstrate plasmon propagation along one of the nanowires by local optical excitation on one end of the wire and collection of fluorescence photons at the opposite end. This is done by focusing $80~\mu W$ of the excitation beam (532~nm) polarized perpendicular to the long axis of the wire with the microscope objective onto the end while scanning the image plane containing the wire with a galvanometric mirror. The focused laser beam creates fluorescence, a fraction of which is coupled to the plasmonic mode which then propagates to the distal end of the wires. Due to a symmetry break at the end, a fraction of the plasmonic mode is coupled to the far field which is then collected by the microscope. Figure~\ref{optical_characterization}(d) and ~\ref{optical_characterization}(e) show such  fluorescence images when the laser was focused on spots \lq A\rq ~and \lq B\rq  of Fig. ~\ref{optical_characterization}(c), respectively, and it is clearly seen that plasmons are supported by the wires as they lit up at the end opposite to the excitation end (corresponding to spots \lq B\rq ~and \lq A\rq ~in Fig.~\ref{optical_characterization}(d) and~\ref{optical_characterization}(e), respectively). We also note that part of the 532~nm light can also get coupled to the plasmonic mode and, in principle, propagate to the distal end where silver nanoclusters can be excited and subsequently fluoresce~\cite{Remoteexcitation}. However, since the estimated loss for 532~nm light for reaching the distal end is around 95~dB, we conclude that this process is unlikely to take place in practice.

At the distal end of the nanowire some of the fluorescence light is reflected back into the plasmonic mode, and thus the nanowire works as a cavity. This can be proven by measuring the spectrum of the fluorescence, which in case of a cavity effect should exhibit oscillations with a period equal to the free-spectral-range of the cavity. Spectra of the wire in Fig.~\ref{optical_characterization}(c) taken at both spots \lq A\rq ~and \lq B\rq ~are presented in Fig.~\ref{optical_characterization}(f). The spectrum of spot \lq A\rq ~in Fig.~\ref{optical_characterization}(f) originates mainly from the fluorescence light that is not coupled into the wire and thus it does not show oscillations but purely the spectrum of the fluorescence. However, the spectrum obtained from spot \lq B\rq ~(the fluoresence, which purely comes from the propagating plasmon) exhibits strongly damped oscillations thereby proving the presence of a very low-finesse cavity.

\section{Simulations and discussion}

\begin{figure}[htbp]
\begin{center}
\includegraphics[]{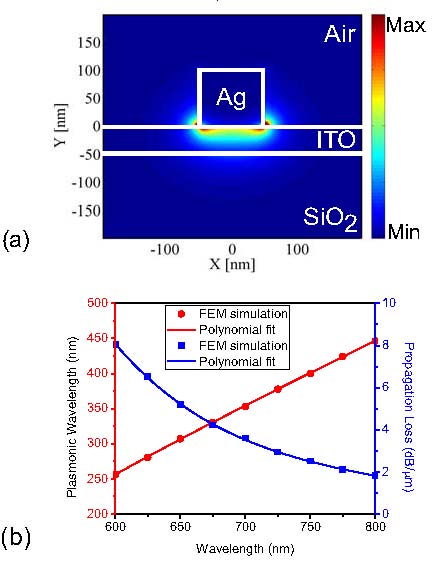}
\caption[FEM simulation of silver nanowires obtained with FIB milling of silver nanoplates.]{\textbf{FEM simulations.} \textbf{(a)} The z-component of Poynting vectors for the fundamental plasmonic mode at a vacuum wavelength of 700~nm is plotted for a silver wire of cross-section 100~nm by 100~nm situated on ITO coated fused silica substrate. \textbf{(b)} Plasmon wavelength and propagation losses as a function of vacuum wavelength. \label{FIB:Simulations}}
\end{center}
\end{figure}

We calculate the fundamental plasmonic mode of a silver wire using a Finite Element Method (FEM) with a commercial software (COMSOL). The Poynting vector ($P_z$) for the fundamental plasmonic mode at a vacuum wavelength of 700~nm of a wire (with a $100 \times 100~{nm}^2$ cross-section) is shown in Fig.~\ref{FIB:Simulations}(a). One clearly sees that the mode is strongly confined at the edges of the wire near the interface to the ITO layer. We also obtained different effective refractive indices for a range of vacuum wavelengths as electric permittivity of silver and ITO varies between  600~nm and 800~nm~\cite{Palik, ITO}. From the simulations we find the plasmon propagation losses ($\alpha_{pl}$) as well as the plasmonic wavelength ($\lambda_{pl}$). The results are presented in Fig.~\ref{FIB:Simulations}(b) as a function of the vacuum wavelength and the data points were fitted to a polynomial of fourth order. The large propagation losses are a result of the strong localization at the ITO-silver interface.

Using the results from the FEM simulation and the fluorescence spectrum from wire end \lq A\rq ~in Fig.~\ref{optical_characterization}(f), the expected spectrum for the oscillating plasmonic field is also shown in Fig.~\ref{optical_characterization}(f) by the solid curve. It shows modulation features similar to that of the spectrum obtained experimentally. In the calculations we have not taken into account the fact that the coupling of fluorescence light into the wire as well as the reflections of the plasmons at the wire ends depend on the wavelength~\cite{reflectionwireend}. Moreover, the cross-section of the fabricated nanowire may not match exactly to the cross-section of the simulated wire. These facts might explain the discrepancy between the  experimentally obtained spectrum and the simulated one. However, the relatively close match between theoretical and experimental spectrum suggests that the wires fabricated are of high quality.

\section{Conclusion}

We have made single crystalline and very smooth silver plates with a thickness around 65~nm and lengths and widths of the order of $10~\mu m$ using a wet chemical synthesis. These plates were then used for the fabrication of nanowires and nanotips with very high precision using FIB milling. The structures were optically addressed, and it was found that the wires support propagating plasmons and that they act as low-finesse cavities. The optical characterization also revealed the presence of large fluorescence which probably stems from redeposition of small silver particles during the FIB milling process.

\section*{Acknowledgments}

The authors acknowledge financial support by the Villum Kann Rasmussen foundation, the Carlsberg foundation, and the Danish council for independent research - natural sciences (FNU). We thank Dr. Jian-Li Mi for assistance with the XRD measurement.


\begin{thebibliography}{99}

\bibitem{2010Bozhevolnyi}
D. K. Gramotnev, and S. I. Bozhevolnyi, \lq\lq Plasmonics beyond the diffraction limit,\rq\rq
\newblock {Nat. Photon.} \textbf{4}, 83--91 (2010).

\bibitem{2010Schuller}
J. A. Schuller, E. S. Barnard, W. Cai, Y. C. Jun, J. S. White, and M. L. Brongersma, \lq\lq Plasmonics for extreme light concentration and manipulation,\rq\rq
\newblock {Nat. Mater.} \textbf{9}, 193--204 (2010).

\bibitem{2006Chang}
D. E. Chang, A. S. S\o{}rensen, P. R. Hemmer, and M. D. Lukin, \lq\lq Quantum optics with surface plasmons,\rq\rq
\newblock {Phys. Rev. Lett.} \textbf{97} (5), 053002 (2006).

\bibitem{2007Chang}
D. E. Chang, A. S. S\o{}rensen, E. A. Demler, and M. D. Lukin,  \lq\lq A single-photon transistor using nanoscale surface plasmons, \rq\rq
\newblock {Nat. Phys.} \textbf{3}, 807--812 (2007).

\bibitem{PRLAgCavity}
H. Ditlbacher, A. Hohenau, D. Wagner, U. Kreibig, M. Rogers, F. Hofer, F. R. Aussenegg, and J. R. Krenn,  \lq\lq Silver nanowires as surface plasmon resonators,\rq\rq
\newblock {Phys. Rev. Lett.} \textbf{95}, 257403 (2005).

\bibitem{2010Hecht}
J. S. Huang, V. Callegari, P. Geisler, C. Br\"{u}ning, J. Kern, J. C. Prangsma, X. Wu, T. Feichtner, J. Ziegler, P. Weinmann, M. Kamp, A. Forchel, P. Biagioni, U. Sennhauser, and  B. Hecht,  \lq\lq Atomically flat single-crystalline gold nanostructures for plasmonic nanocircuitry.\rq\rq \newblock {Nat. Commun.} \textbf{1}, 350--357 (2010).

\bibitem{Auwirefabrication}
F. Kim, K. Sohn, J. Wu, and J. Huang,  \lq\lq Chemical synthesis of gold nanowires in acidic solutions, \rq\rq
\newblock {J. Am. Chem. Soc.} \textbf{130}(44), 14442--14443 (2008).

\bibitem{Agwirefabrication}
K. E. Korte, S. E. Skrabalak, and Y. Xia,  \lq\lq Rapid synthesis of silver nanowires through a cucl- or cucl2-mediated
  polyol process, \rq\rq
\newblock {J. Mater. Chem.} \textbf{18}, 437--441 (2008).

\bibitem{2007Akimov}
A. V. Akimov, A. Mukherjee, C. L. Yu, D. E. Chang, A. S. Zibrov, P. R. Hemmer,
  H. Park, and M. D. Lukin,  \lq\lq Generation of single optical plasmons in metallic nanowires coupled to quantum dots, \rq\rq
\newblock {Nature} \textbf{450} (06230), 402--406 (2007).

\bibitem{2009Kolesov}
R. Kolesov, B. Grotz, G. Balasubramanian, R. J. St\"{o}hr, A. A. L. Nicolet, P. R. Hemmer, F. Jelezko, and
  J. Wrachtrup,  \lq\lq Wave-particle duality of single surface plasmon polaritons,\rq\rq
\newblock {Nat. Phys.} \textbf{5}, 470--474 (2009).

\bibitem{2011Huck}
A. Huck, S. Kumar, A. Shakoor, and U. L. Andersen,  \lq\lq Controlled coupling of a single nitrogen-vacancy center to a silver
  nanowire,\rq\rq
\newblock {Phys. Rev. Lett.} \textbf{106} (8), 096801 (2011).

\bibitem{Auplatefabrication}
Z. Guo, Y. Zhang, Y. D. Mu, L. Xu, S. Xie, and N. Gu,  \lq\lq Facile synthesis of micrometer-sized gold nanoplates through an
  aniline-assisted route in ethylene glycol solution,\rq\rq
\newblock {Colloid Surf. A-Physicochem. Eng. Asp.} \textbf{278} (1--3), 33--38 (2006).

\bibitem{MicrosizedAgcrystals}
C. Xionghui,  and Z. Aixia,  \lq\lq Preparation of microsized silver crystals with different morphologies
  by a wet-chemical method,\rq\rq
\newblock {Rare Metals} \textbf{29}, 407--412 (2010).

\bibitem{Refereedemand}
C. Gruber, P. Kusar, A. Hohenau, and J. R. Krenn, \lq\lq Controlled addressing of quantum dots by nanowire plasmons,\rq\rq Appl. Phys. Lett. \textbf{100}, 231102 (2012).

\bibitem{Goldflakemachine}
C. S. Ah, Y. J. Yun, H. J. Park, W.-J. Kim, D. H. Ha, and W. S. Yun,  \lq\lq Size-controlled synthesis of machinable single crystalline gold nanoplates,\rq\rq
\newblock {Chem. Mat.} \textbf{17} (22), 5558--5561 (2005).

\bibitem{JohnsonChristy}
P. B. Johnson, and R. W. Christy,  \lq\lq Optical constants of the noble metals,\rq\rq
\newblock {Phys. Rev. B} \textbf{6}, 4370--4379 (1972).

\bibitem{Lowlossmaterial}
P. R. West, S. Ishii, G. V. Naik, N. K. Emani, V. M. Shalaev, and A. Boltasseva,  \lq\lq Searching for better plasmonic materials,\rq\rq
\newblock {Laser Photon. Rev.} \textbf{4} (6), 795--808 (2010).

\bibitem{Liquidliquidinterface}
M.-S. Jin, Q. Kuang, X.-G. Han, S.-F. Xie, Z.-X. Xie, and L.-S. Zheng,  \lq\lq Liquid–liquid interface assisted synthesis of size- and
  thickness-controlled ag nanoplates,\rq\rq
\newblock {J. Solid State Chem.} \textbf{183} (6), 1354--1358 (2010).

\bibitem{langmuir}
T. Deckert-Gaudig, F. Erver, and V. Deckert,  \lq\lq Transparent silver microcrystals: Synthesis and application for
  nanoscale analysis,\rq\rq
\newblock {Langmuir}, \textbf{25} (11), 6032--6034 (2009).

\bibitem{solvothermal}
Q. Lu, K.-J. Lee, S.-J. Hong, N. V. Myung, H.-T. Kim, and Y.-H. Choa,  \lq\lq Growth factors for silver nanoplates formed in a simple solvothermal
  process,\rq\rq
\newblock {J. Nanosci. Nanotechnol.} \textbf{10} (5), 3393--3396 (2010).

\bibitem{FIBbook}
L. A. Giannuzzi and F. A. Stevie, {\em Introduction to Focused Ion Beams: Instrumentation, Theory,
  Techniques and Practice}
\newblock (Springer, USA, 2005).

\bibitem{Agfluorescence}
L. A. Peyser, A. E. Vinson, A. P. Bartko, and R. M. Dickson,  \lq\lq Photoactivated fluorescence from individual silver nanoclusters,\rq\rq
\newblock {Science} \textbf{291} (5501), 103--106 (2001).

\bibitem{Remoteexcitation}
H. Wei, D. Ratchford, X. Li, H. Xu, and C.-K. Shih,  \lq\lq Propagating surface plasmon induced photon emission from quantum
  dots.\rq\rq
\newblock {Nano Lett.} \textbf{9} (12), 4168--4171 (2009).

\bibitem{Palik}
E. D. Palik,  {\em Handbook of Optical Constants}
\newblock (Academic Press, 1998).

\bibitem{ITO}
S. Laux, N. Kaiser, A. Z\"{o}ller, R. G\"{o}tzelmann, H. Lauth, and H. Bernitzki,  \lq\lq Room-temperature deposition of indium tin oxide thin films with plasma ion-assisted evaporation,\rq\rq
\newblock {Thin Solid Films} \textbf{335} (1–2), 1--5 (1998).

\bibitem{reflectionwireend}
R. Gordon,  \lq\lq Reflection of cylindrical surface waves,\rq\rq
\newblock {Opt. Express} \textbf{17} (21), 18621--18629 (2009).


\end{thebibliography}
\end{document}